\documentclass[12pt]{article}
\usepackage{graphicx,amsmath}
\usepackage{units}
\usepackage{cite}
\usepackage{amssymb}

\newlength{\dinwidth}
\newlength{\dinmargin}
\def\lapproxeq{\lower .7ex\hbox{$\;\stackrel{\textstyle                                                    
<}{\sim}\;$}}                                                    
\def\gapproxeq{\lower .7ex\hbox{$\;\stackrel{\textstyle                                                    
>}{\sim}\;$}}                                                    
\def\be{\begin{equation}}                                                    
\def\ee{\end{equation}}                                                    
\def\bea{\begin{eqnarray}}                                                    
\def\eea{\end{eqnarray}}

\def\sh{\hat s}
\def\sh2{{\hat s}^2}

\begin{document}
                                                    
\titlepage                                                    
\begin{flushright}                                                    
IPPP/18/71  \\                                                    
\today \\                                                    
\end{flushright} 
\vspace*{0.5cm}
\begin{center} 
{\Large\bf Searching for the Odderon in Ultraperipheral \\ \vspace{0.3cm} Proton--Ion Collisions at the LHC}\\

\vspace*{1cm}
                                                   
L.A. Harland-Lang$^a$, V.A. Khoze$^{b,c}$, A.D. Martin$^b$ and M.G. Ryskin$^{b,c}$ \\                                                    
                                                   
\vspace*{0.5cm}                                                    
$^a$ Rudolf Peierls Centre, Beecroft Building, Parks Road, Oxford, OX1 3PU\\
$^b$ Institute for Particle Physics Phenomenology, University of Durham, Durham, DH1 3LE \\                                                   
$^c$ Petersburg Nuclear Physics Institute, NRC Kurchatov Institute, Gatchina, St.~Petersburg, 188300, Russia

\vspace*{1cm}                                                    
                                                    
\begin{abstract}          

\noindent
We explore the possibility of observing Odderon exchange in proton--ion collisions at the LHC, via the ultraperipheral photoproduction of 
C-even mesons.  As well as the signal, we consider in detail the principle backgrounds, due to QCD--initiated production (i.e. double Pomeron exchange) and $\gamma\gamma$ fusion. We find that while the photon--initiated background is dominant at very small momentum transfer, this can be effectively removed by placing a reasonable cut on the transverse momentum of the produced meson. In the case of QCD--initiated production, we show this is in general strongly suppressed by the small probability of no additional particle production in the central detector, namely the survival factor. In some scenarios, this suppression is sufficient to permit the observation of Odderon exchange in Pb$-p$ collisions in a clean environment, or else to place bounds on this. We in addition identify the cases of $\pi^0$ and $\eta(548)$ production as particularly promising channels. Here, the QCD--initiated background is absent for $\pi^0$ due to isospin conservation and very small for $\eta(548)$ due to its dominantly flavour octet nature and odd parity.
 
\end{abstract}                                                        
\vspace*{0.5cm}                                                    
                                                    
\end{center}

 \section{Introduction}

The TOTEM experiment at the LHC \cite{Antchev:2017yns}
has recently published the
results of the first high precision measurement at  $\sqrt s=$13 TeV
of the ratio of
the real-to-imaginary parts  of the forward elastic 
$pp$-amplitude,  $\rho=$Re$A$/Im$A$.
The measured value of $\rho=0.09-0.10$  is smaller
than that predicted by the
conventional COMPETE parametrization ~\cite{Patrignani:2016xqp}, which gives $\rho=0.135$. 
This may indicate either a slower increase of the total cross section at  higher energies or
a  manifestation of the odd-signature amplitude, which is not included in the COMPETE parametrization.

This TOTEM result has generated renewed interest~\cite{Martynov:2017zjz,Khoze:2018bus,
Khoze:2017swe,Goncalves:2018yxc,Goncalves:2018pbr,Martynov:2018nyb,Khoze:2018kna,
Shabelski:2018jfq,Broilo:2018brv,Broilo:2018els,Troshin:2018ihb,Broniowski:2018xbg,
Csorgo:2018uyp,Martynov:2018sga,
Bence:2018ain,Dremin:2018uwt,Selyugin:2018uob}
in the long--standing issue of establishing the 
existence
 of the odd--signature partner of the
Pomeron, the so--called Odderon. This was first introduced in the early seventies
 in the framework of asymptotic theories~\cite{Lukaszuk:1973nt,Joynson:1975az},
and since then has been the subject of intensive theoretical discussion, see for
 example~\cite{Braun:1998fs,Ewerz:2003xi,Ewerz, 
Ewerz:2009zz} for reviews. This  odd--signature exchange, which depends only weakly on energy, is a firm prediction of QCD~\cite{Kwiecinski:1980wb,Bartels:1980pe} (see e.g.~\cite{Ioffe:2010zz} for a more recent textbook discussion). In~\cite{Bartels:1999yt} it was shown that in the perturbative regime there exists
 a colourless C--odd $t$-channel state, formed by  three gluons, with an intercept $\alpha_{\rm Odd}=1$.  
 
 In this paper we consider the possibility of observing Odderon exchange at the LHC via the semi-exclusive production of a C-even meson $M$ in ultraperipheral heavy-ion-proton collisions, that is via the subprocess
 \be
 \gamma + p \to M\,+\,X\;,
 \ee
 where $X$ denotes the dissociation product of the proton, and the photon is emitted elastically from the ion. This subprocess is shown in Fig.~\ref{fig:1}(a), while the corresponding   production process in $pPb$ collisions is  shown in Fig.~\ref{fig:1}(b), see~\cite{Goncalves:2018yxc} for recent work. In this case the signal is significantly enhanced by the photon flux from the ion by $\sim Z^2$, and as we will see the expected cross sections are certainly within reach at the LHC. However, for a full analysis it is essential to provide a proper consideration of the potential background contributions to this process, due either to double Pomeron exchange, or $\gamma\gamma$ fusion. In reality, it is the magnitude of these backgrounds, rather than necessarily the signal size, that determines the level at which we could expect to observe or put new limits on the Odderon contribution. We therefore present in this paper the first complete calculation of both the expected Odderon signal and background contributions for $C$ even meson production at the LHC. As we shall discuss further below, an evaluation of the double pomeron exchange background requires a careful treatment of the proton--ion or ion--ion collision process. We will see that,  while in ion--ion collisions the background from $\gamma\gamma$ is overwhelming, in proton--ion collisions it is expected to be under better control. 

The outline of this paper is as follows. In Section~\ref{sec:oddsearch} we briefly summarise the current experimental situation with regards to Odderon exchange, with emphasis on the phenomenology of $C$--even meson photoproduction. In Section~\ref{sec:gen} we present some general considerations for the conversion of the $\gamma p$ subprocess cross section to the case of $pPb$ collisions, and translate the existing HERA results to expected upper limits on the Odderon signal for $\pi^0$, $f_2$ and $\eta$ production at the LHC. In Section~\ref{sec:oddcalc} we present a phenomenological calculation of meson production via Odderon--exchange, concentrating on the $\pi^0$ case for concreteness. In Section~\ref{sec:bggam} we discuss and provide a quantitative calculation of the background from $\gamma\gamma$ fusion. In Section~\ref{sec:bgpom} we discuss and provide a quantitative calculation of the background from double pomeron exchange.  In Section~\ref{sec:bgrad} we discuss the reducible background arising from the radiative decay of vector mesons. In Section~\ref{sec:num} we present a detailed numerical analysis of the signal and background cross sections for the cases of $f_2$, $\pi^0$, $\eta$ and $\eta_c$ production. Finally, in Section~\ref{sec:conc} we conclude.

\section{Searching for evidence of Odderon exchange}\label{sec:oddsearch}

Although in QCD we expect both Pomeron and Odderon exchange to contribute to scattering processes,  the  pre--LHC experimental quest for the Odderon  has proven to be quite a challenging task, see for instance~\cite{Braun:1998fs, Ewerz:2009zz, Ewerz,Block} for reviews.
In particular, the contribution from Odderon exchange  to elastic $pp$-scattering is predicted to be
rather small, see e.g. \cite{Ryskin:1987ya,Levin:1990gg}, providing only a  small correction to the dominant even-signature Pomeron exchange. Moreover, due to screening effects this contribution is expected to decrease with increasing energy, see \cite{Khoze:2018bus,Ryskin:1987ya,Levin:1990gg,Finkelstein:1989mf}.

\begin{figure} 
\begin{center}
\includegraphics[scale=0.6]{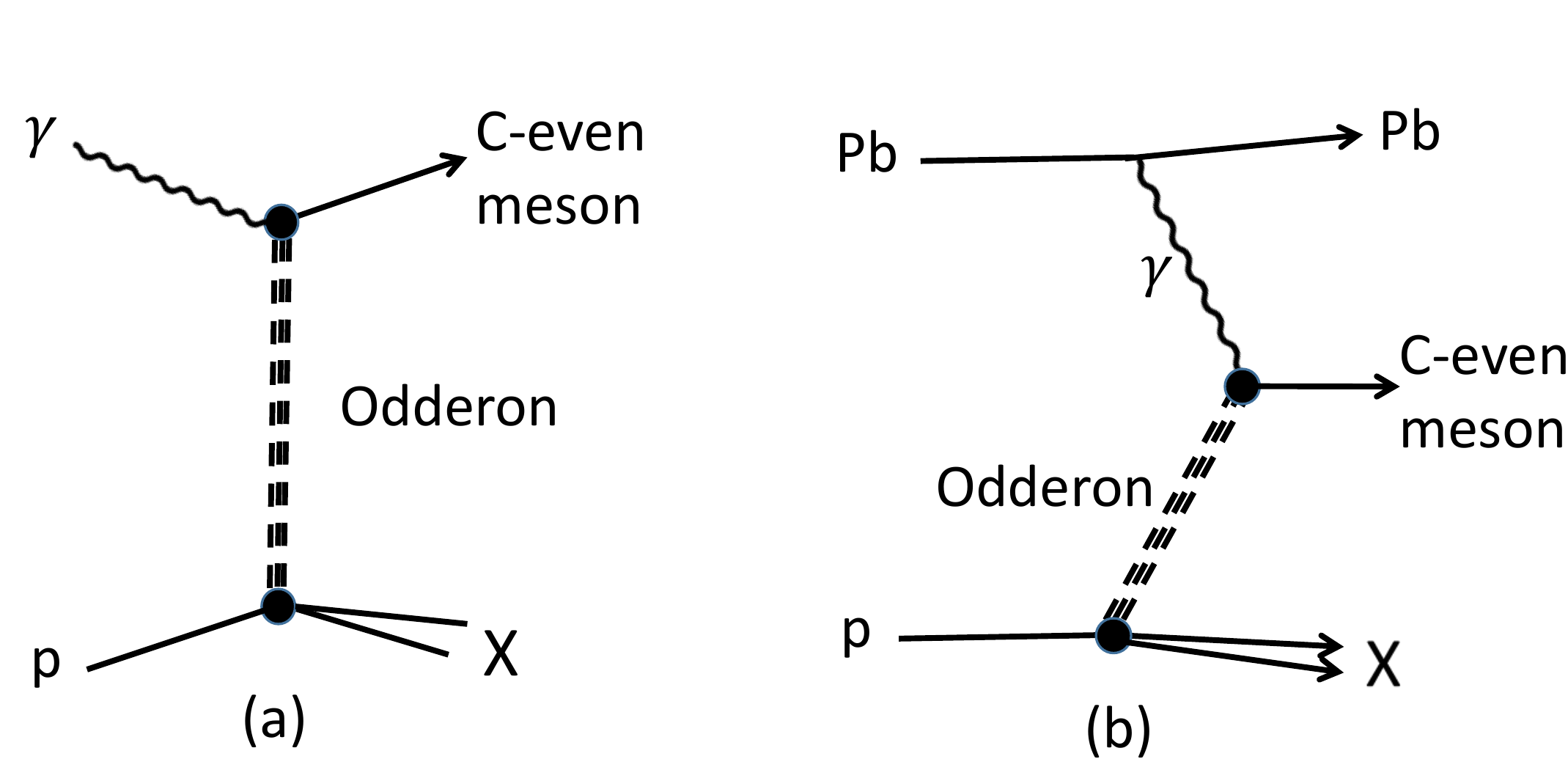}
\caption{\sf The Odderon exchange contribution to (a) photoproduction of C-even mesons and (b) exclusive ultraperipheral C-even meson production in Pb$-p$ collisions.}
\label{fig:1}
\end{center}
\end{figure}

The possibility of probing the Odderon via the high energy exclusive photoproduction of C-even mesons, 
which can be mediated by this odd-signature exchange, has a long history, see e.g.~\cite{Barakhovsky:1991ra,Ryskin:1998kt,Kilian:1997ew,Rueter:1998gj,Berger:1999ca,
Berger:2000wt,Ivanov:2001zc,Bzdak:2007cz}.
In particular, above the very low transverse momenta region where the major contribution comes from
photon exchange, the Odderon may dominate, see Fig.~\ref{fig:1} (a). The cross sections for light $C$--even meson ($\pi^0,\eta,f_2)$  photoproduction were evaluated for example in~\cite{Ryskin:1998kt,Rueter:1998gj,Berger:2000wt}, by applying a pQCD treatment with an effective cutoff applied to regulate the contribution from the infrared region.

The expected rates become rather large when break up of the target proton is permitted.
In particular, an Odderon-induced cross section of $\sim$ 300 nb has been predicted for the
 $\gamma p\to\pi^0 +X$ reaction~\cite{Rueter:1998gj},  and about 21 nb for the  $\gamma p\to f_2 +X$ case
\cite{Berger:2000wt}, at  $\sqrt{s}=20$ GeV.
 Since the $\eta$ meson belongs to the same multiplet as the pion we may expect more or less the same cross section for 
  $\gamma p\to\eta+X$ photoproduction. More precisely, the rate will be a little lower due to the branching ratio of the considered $\eta$-decay and the smaller  $u$-quark fraction in the wave function of the $\eta$ meson.
  
Searches for the Odderon in the 
high energy photoproduction of C-even mesons were performed at HERA at $\sqrt 
s\simeq 200$ GeV~\cite{Olsson:2001nm,Adloff:2002dw,Berndt:2002tw}. No signal was observed, and upper limits of 39 nb for  
$\pi^0$ and  16 nb for $f_2$ production, were set  at the 95\% confidence level, in conflict with these higher
predictions. 
However,
 the results in \cite{Berger:1999ca,Berger:2000wt}
are only based on a leading order calculation in $\alpha_S$. Moreover, 
the major contribution in the corresponding integrals comes from the infrared domain (especially for the $\pi^0$), 
where it should be suppressed by confinement.
In addition, the results are highly sensitive to the details of the proton wave function. For example, in the extreme limit of a proton formed by a quark and a point-like diquark, the Odderon to proton coupling, $g_{\rm{O}p}$,  in fact vanishes (see e.g. ~\cite{Ryskin:1987ya,Levin:1990gg, Rueter:1996yb}).

Away from this limit, the Odderon--proton coupling is nonetheless expected to be small. This in fact highlights one of the general advantages of using inelastic photoproduction to search for Odderon exchange, namely that the corresponding amplitude is proportional to the first power of the small Odderon coupling $g_{\rm{O}p}$, while in elastic $pp$-scattering the amplitude is proportional to $g^2_{\rm{O}p}$. Moreover, by selecting events with larger meson transverse momenta, $k_t$, we can limit the sensitivity to the non--perturbative regime and corresponding infrared cutoff. 
As discussed in detail in Section~\ref{sec:oddcalc}, a pQCD based calculation suggests that a lower, but still sufficiently large cross section of
\be
\sigma(\gamma+p\to M+X) \sim 0.5-5~ {\rm nb}\;,
\ee
is reasonable.  As a result in  $p$Pb collisions we expect a cross section 
\be
\sigma(p+\mbox{Pb}\to X+M+\mbox{Pb})~ \sim ~ 0.1-1~\mu{\rm b}\;,
\ee
for Odderon-mediated photoproduction at the LHC via the semi-exclusive central process, which we will denote as CEP* in what follows.
We emphasise that these estimates will serve as guidance when assessing the potential for observing Odderon exchange at the LHC. However, in reality, the HERA upper limits on the photoproduction cross sections will  provide the clearest indication of the possibility for $p$Pb collisions at the LHC to observe the Odderon.  The emphasis below will be on calculating the size of the other contributing backgrounds, as is mandatory in a complete assessment of the LHC discovery potential.

\section{Odderon exchange in $p$Pb collisions: general considerations}\label{sec:gen}

The Odderon--exchange cross section for the production of a meson $M$ in $p$Pb collisions can we written as
\be
\frac{{\rm d}\sigma^{p\rm Pb}}{{\rm d} Y_M} \sim \frac{{\rm d}N_{\gamma}}{{\rm d} Y_M}\sigma^{\gamma p}\;,
\ee
where $N_\gamma$ is the photon flux due to the lead ion and $Y_M$ is the meson rapidity. Here, we have used the fact that the cross section due to Odderon exchange is expected to be independent of the  $\gamma p$ energy, and neglected the rather smaller impact of the $pPb$ survival factor. The advantage of using heavy ions is the significant enhancement of the photon flux; for lead the coherent $\gamma$-flux (i.e. with the lead ion remaining intact) is enhanced by $Z^2=82^2$. More precisely, the $\gamma$-flux reads (see e.g. \cite{Fermi:1924tc,Landau:1934zj,Budnev:1974de}) 
\be 
\frac{{\rm d}N_{\gamma}}{{\rm d}Y_M}\simeq \frac{\alpha_{\rm QED}}\pi\int \frac{q^2_\perp}{(q^2_\perp+(xm_p)^2)^2}F_Z(Q^2)^2\,{\rm d}q^2_\perp\;,
\ee 
where $m_p$ is the proton mass, $F_Z(Q^2)$ is the ion form factor, and $x$ is the momentum fraction carried by the photon. As an example, if we consider $\eta$ production in $p$Pb collisions at $\sqrt s_{NN}=8.8$ TeV (corresponding to the maximum collision energy at the LHC), then at central rapidity $Y_M=0$ we have
\be
x\sim \frac{M_\eta}{\sqrt s}=6.2\cdot 10^{-5}\;,
\ee
and so
\be
\frac{{\rm d}N_{\gamma}}{{\rm d}Y_M}\bigg|_{Y_M=0}\simeq 190\;.
\ee
For the Odderon--exchange cross sections, we therefore simply require the corresponding prediction for the $\gamma p$ cross sections. 
Unfortunately, as mentioned above, the predicted cross section 
is very sensitive to the treatment of the infrared QCD regime. While we will discuss some model--dependent estimates for this below, the most conservative approach is therefore simply to take the HERA 95\% confidence upper limits on the $\pi^0$ and $f_2$  cross sections in $\gamma p$ collisions as guidance.

For the case of  pseudoscalar $\eta$ production, we can use the flavour decomposition of~\cite{Feldmann:1998vh}
\be 
|\eta\rangle = f_8\cos \theta_8 |q\overline{q}_8\rangle - f_1\sin \theta_1|q\overline{q}_1\rangle + |gg\rangle\;,
\ee
with
\be 
|q\overline{q}_1\rangle =\frac{1}{\sqrt{3}}|u\overline{u}+d\overline{d}+s\overline{s}\rangle\;,\qquad
|q\overline{q}_8\rangle =\frac{1}{\sqrt{6}}|u\overline{u}+d\overline{d}-2s\overline{s}\rangle\;,
\ee
and 
\begin{align}\nonumber
  f_8=1.26 f_\pi\;, \qquad & \qquad \theta_8   = -21.2^\circ\;,\\ \label{thetafit}
 f_1=1.17 f_\pi\;,  \qquad & \qquad \theta_1  = -9.2^\circ\;.
\end{align}
While the gluonic component of course does not contribute to photoproduction, the amplitudes for the $q\overline{q}$ contributions are weighted by the quark charges, with
\be
\mathcal{M}_{\gamma p }^1 \propto \frac{1}{\sqrt{3}}\left(\frac{2}{3} - \frac{1}{3} -\frac{1}{3}\right) = 0 \;,\qquad \mathcal{M}_{\gamma p }^8 \propto \frac{1}{\sqrt{6}}\left(\frac{2}{3} - \frac{1}{3} +\frac{2}{3}\right) = \frac{1}{\sqrt{6}}\;.
\ee
For the $\pi^0$ we will pick up a factor of $1/\sqrt{2}$ from the wave function, and hence we have
\be
\frac{\sigma^\eta}{\sigma^{\pi^0}} ~\simeq ~ \frac{1}{3} \frac{f_8^2}{f_\pi^2} \cos^2\theta_8 ~=~ 0.46\;.
\ee
We can then scale the $\pi^0$ HERA limit by this factor to give the corresponding limit for the $\eta$. These results are summarised in Table~\ref{tab:oddcs}.  

\begin{table}
\begin{center}
\begin{tabular}{|c|c|}
\hline
${\rm d}\sigma/{{\rm d} Y_M}|_{Y_M=0}$ &Expected upper limits [$\mu$b]\\
\hline
$\pi^0$&7.4\\ \hline  
$\eta$  &3.4\\  \hline  
$f_2(1270)$  &3.0 \\  \hline    
\end{tabular}
\caption{\sf Expected upper limits for the differential production cross section at $Y_M=0$, of light mesons in $p$Pb collisions, via Odderon--exchange. The values are calculated using the corresponding HERA upper limits~\cite{Olsson:2001nm} on the $\pi^0$ and $f_2$ cross sections in $\gamma p$ collisions, scaled by the ion flux. The $\eta$ limit is calculated from the $\pi^0$ value, and scaled by the expected cross section ratio, as described in the text.} \label{tab:oddcs}
\label{Tab1}
\end{center}
\end{table}

\section{Meson production via Odderon exchange: cross section calculation}\label{sec:oddcalc}

In this section we consider the calculation of $\pi^0$ photoproduction for concreteness, but will comment on the cases of $f_2$ and $\eta$ meson production at the end.
To calculate the cross section for $\pi^0$ photoproduction via Odderon exchange it is most convenient to relate it to the cross section due to $\gamma\gamma$ fusion; we will discuss this background further in the following section. We have
\be\label{eq:dsigodd}
\frac{d\sigma^{\gamma p}_{\rm Odd}}{dt}= \left(\frac{9}{5}\right)^2 \cdot 3\cdot 
\frac{|T_{\rm Odd}(t)|^2}{|T_{\rm QED}(t)|^2}\cdot \frac{d\sigma^{\gamma p}_{\rm QED}}{dt}\;.
\ee
The origin of the various terms is as follows. First, $T_{\rm Odd}$ ($T_{\rm QED}$) are the basic quark--quark scattering amplitudes via Odderon (photon) exchange, with the quark electric charge factored out in the latter case. Then, the factor of $(9/5)^2$ accounts for the electric charge weighting of the quarks within the pion, absent in the Odderon case. Next, on the proton side, at  relatively large $|t|$, where the dominant contribution comes from  proton ($p\to X$) dissociation, we can roughly speaking consider our scattering process as being due to 3 independent interactions with the individual valence quarks in the proton. This gives $\sum e^2_i=1$ for the QED cross section but the number of valence quarks, i.e. a factor 3 for the Odderon cross section\footnote{More precisely, in the case of photon exchange the inelastic photon flux from the proton is well known~\cite{Manohar:2016nzj}. However for the purposes of estimating the Odderon cross section, this approximation is sufficient, while for the photon initiated background, as we will see, the suppression after imposing suitable cuts is large enough that a more precise calculation is not necessary.}. Finally, we must consider the diagrams where the three gluons comprising the Odderon couple not only to the same quark but to two quarks or each gluon to its own valence quark. 
This contribution contains the interference between different diagrams and strongly depends on the details of proton wave function. Neglecting interference (which is justified at large $|t|$) we expect in (\ref{eq:dsigodd}) an additional factor of 8. However, the $|t|$ values we will consider are not so large, and so to be conservative we only account here for the diagram where all three gluons couple to the same quark line.

To evaluate the expected cross section due to Odderon exchange we then require the $\pi^0$ photoproduction cross section, including proton dissociation. This can be calculated using the known $\pi^0\to\gamma\gamma$ width, to give
\begin{equation}
\label{pi0}
\frac{d\sigma^{\gamma p}_{\rm QED}}{dt}=\frac{4\pi\Gamma_{\gamma\gamma}~}{m^3}\frac{\alpha}{|t|}F^2_\pi(t)F^2_p(t)~=~\frac{0.11\mbox{nb}}{|t|} F^2_\pi(t)\;,
\end{equation}
where in the last equality we have neglected the proton form factor $F_p(t)$, and allow the proton to dissociate. For the $t$-dependence of the pion form factor we apply the simplified $\rho$ meson pole formula
\begin{equation}
\label{f-pi}
F_\pi(t)=\frac 1{1-t/0.6 ~\mbox{GeV}^2}\;.
\end{equation}
For the Odderon exchange amplitude we have to account for the permutations of gluons in the pion-Odderon vertex, that is include the diagrams where two gluons couple  to the  quark and the third gluon couples to the antiquark, and vice-versa.
Thus $T_{\rm Odd}(t)$ takes the form
\be
\label{o1}
 T_{\rm Odd}(t)=\frac{10\alpha_s^3(q^2)}{81\pi}\int\left[F_\pi(q^2)-F_\pi(4(q/2-q_1)^2)\right]\frac{d^2q_1d^2q_2d^2q_3\delta^{(2)}(q-q_1-q_2-q_3)}{(q^2_1+\mu^2)(q^2_2+\mu^2)(q^2_3+\mu^3)}\;,
\ee
where  $q_i$ is the transverse momentum of the $t$-channel gluon $i$ and $q^2=t$. The first factor accounts for the colour coefficients, the identity of the 3 gluons and factors (such as $2\pi$) corresponding to the Feynman loop integration. To regulate the infrared divergence associated with the loop integral we have introduced an artificial cutoff $\mu$ in the gluon propagators. The first term in the square brackets corresponds to the diagram where all three gluons couple to the same quark/antiquark, while the second term is responsible for the permutations, see for example~\cite{Levin:1981rf}; due to the symmetry between the 3 gluons it is sufficient to consider only the gluon $q_1$ here. We can see that in the second $F_\pi$ term there is no form factor suppression when $q_1=q/2$. In this case, both non-relativistic quarks have half of the momentum transfer $q$, and thus go in the same direction.

Finally, to convert this to a corresponding cross section using \eqref{eq:dsigodd} we have to multiply the amplitude $T_{\rm QED}$, which describes the QED interaction between two quarks (with the electric charges $e_q=1$), by the form factor $F_\pi$
 \be\label{e1}
  T_{\rm QED}=\frac{8\pi \alpha}{t}F_\pi(t)\; ,
\ee
 that is we have to use in (\ref{eq:dsigodd}) the amplitudes (\ref{o1}) and (\ref{e1}).

\begin{figure} 
\begin{center}
\vspace*{-1.0cm}
\includegraphics[scale=0.5]{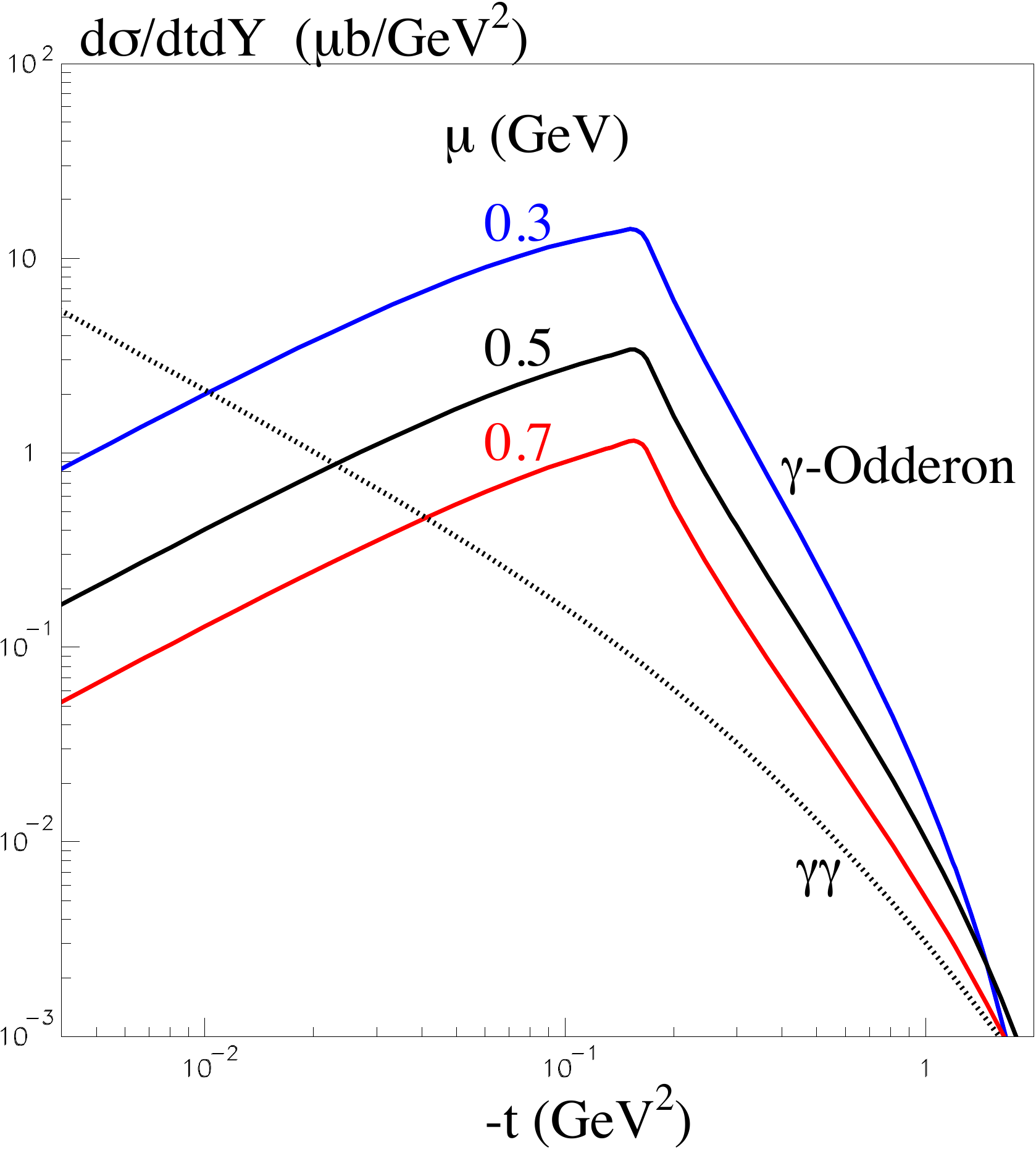}
\caption{\sf Differential cross section for $\pi^0$ central production in proton-lead collisions due to photon-Odderon (solid lines) and photon-photon (dashed line) fusion, where $\sqrt s_{nn}=8.16$ TeV and $Y_M=0$.}
\label{fig:3a}
\end{center}
\end{figure}

 The expected cross sections are shown in Fig.~\ref{fig:3a}. We can see that while at low $t$ the QED cross section is dominant, as $t$ increases the relative contribution from Odderon exchange increases. In particular, for $|t|\gtrsim 0.04$ GeV$^2$, i.e.  $p_\perp>0.2$ GeV, the Odderon cross section is completely dominant. Note that the average transverse momentum of the photon emitted by the ion, $p_\perp \sim 1/R_A $, where $R_A$ is the ion radius, is roughly an order of magnitude smaller than this. Thus we can effectively select this larger $|t|>0.04$ GeV$^2$ region by placing a cut $p_\perp>0.2$ GeV on the produced meson. Other features of note in the figure are the rapid fall off of the Odderon induced cross section as $|t|$ becomes too large, due to the large power of the running coupling $\sim \alpha_s^6$, and by destructive interference between the first and the second terms in square brackets of (\ref{o1}). In addition, the break at $|t|\simeq 0.15$ GeV$^2$ is due to fact that for a smaller $|t|$ the QCD coupling was frozen at $\alpha_s=1$. 
 Integrating over the range 0.04 $<|t|<$ 1 GeV$^2$  (i.e. over the $0.2<p_\perp<1$ GeV interval) we find 
 \be
\sigma^{\gamma p}_{\rm Odd} = 1 - 12\, {\rm nb}\;,
 \ee
 for $\mu = 0.7-0.3$ GeV.

We note that as there is no complete theoretical description of the confinement region of QCD, the above calculation only corresponds to one possibility for dealing with the infrared contribution from \eqref{o1}.
To give an idea of the model dependence in this region, we have also examined an alternative method for dealing with the low $q_{i\perp}$ region. In particular, we explicitly limit the integral to the $q_i>q_0$ region, and in addition we replace the QCD coupling factor $\alpha_s(t)^3$ by a product of couplings evaluated at each $q_i^2$.
 In this case when integrating over the same $0.04<|t|<1$ GeV$^2$ interval we find a smaller cross section of 
 \be
 \sigma^{\gamma p}_{\rm Odd} = 0.07- 1.6 \, {\rm nb}\;,
 \ee
for $q_0=0.5- 0.3$ GeV. Interestingly in this case the relative contribution from the higher $|t| > 1$ GeV$^2$ region is rather large, being of roughly the same size as the above result, e.g. $\sim 1.6$ nb for $q_0=0.3$ GeV. For the QED cross section the contribution from this large $|t|$ region is significantly lower, $\sim 0.01$ nb.
 
 We must also consider the impact of the gap survival factor, $S^2$. As discussed in Section~\ref{sec:gen}, the impact of additional proton--ion interactions is rather small for the ultraperipheral collision process we are considering, and thus $S^2_{pPb}$ is quite close to unity. However, we must also consider the possibility of additional inelastic photon--proton and pion--proton interactions. In the former case, this effect is due to the additional interaction of a $q\overline{q}$ pair created by the photon with the target proton. Unfortunately, there is no appropriate data to constrain the impact of this, and thus any estimate will be rather model dependent. Taking the $\rho$--meson dominance model of~\cite{Kaidalov:2009fp}, we find $S^2_{\gamma p}\sim 0.3$, and thus we may expect the Odderon signal cross section to be suppressed by a factor of  $\sim 3$ relative to the results quoted above.
 
In summary, from the above considerations we roughly expect a signal cross section of 
 \be
 \sigma^{\gamma p}_{\rm Odd} = 0.5- 5 \, {\rm nb}\;,
 \ee
 for $\pi^0$ production, although in reality the cross section could be smaller than this, depending on the specific treatment of the infrared region. 
 
Finally, for the $f_2$ meson we can apply the same procedure as above, accounting for the appropriate $\gamma\gamma$ width. For the $\eta$ we must apply the corrections as discussed in Section~\ref{sec:gen}. In both cases, we find that the cross section is expected to be roughly half that of the $\pi^0$.

\section{Background from photon-photon fusion}\label{sec:bggam}

The size of the background due to $\gamma\gamma$ fusion can be readily calculated using the procedure described in the previous section. In particular, we have seen from Fig.~\ref{fig:3a} that for meson $p_\perp>0.2$ GeV, this background is expected to be very small. To quantify this, we simply apply \eqref{pi0}, and integrate over 0.04 $<|t|<$ 1 GeV$^2$. We find
\be
 \sigma^{\gamma p}_{\rm QED} = 0.23 \, {\rm nb}\;.
\ee
Therefore, comparing to the estimates of the previous section, the background is expected to be under good control.

 \section{Background from Pomeron-Pomeron fusion}\label{sec:bgpom}

\begin{figure} [htb]
\begin{center}
\includegraphics[scale=0.6]{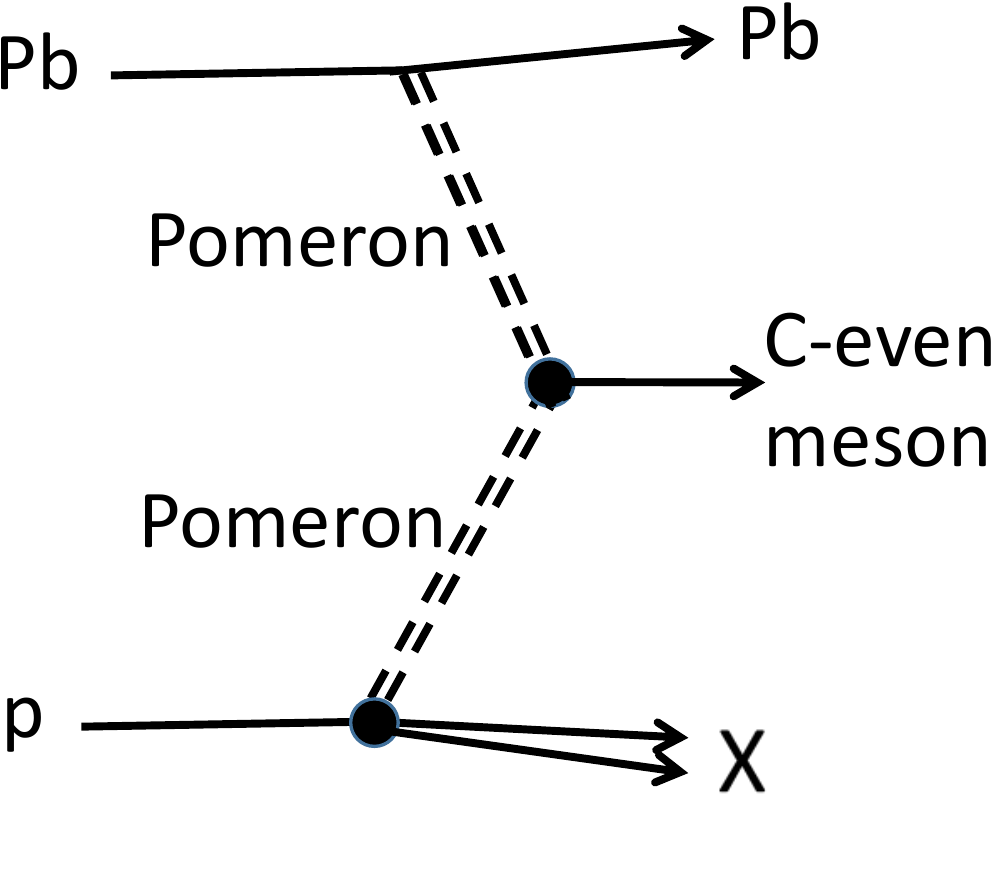}
\caption{\sf A background to the Odderon searches based on Fig.~\ref{fig:1}(b) arising from the production of C-even mesons by Pomeron-Pomeron fusion.}
\label{fig:3}
\end{center}
\end{figure}

In contrast to the odderon searches at HERA via photoproduction, in heavy ion-proton collisions we must also consider the background due to Pomeron--Pomeron fusion, see Fig.~\ref{fig:3}.
While we might naively expect such a QCD--initiated background to be enhanced relative to Odderon--induced photoproduction, in fact for the case of exclusive, or semi--exclusive production (CEP${}^*$), this is strongly suppressed
by the gap survival factor $S^2$.
In particular, in addition to the specific nucleon-nucleon collision which participates in the CEP${}^*$ interaction, there is a significant probability to have additional inelastic interactions between other pairs of nucleons which will populate the rapidity gaps either side of the produced meson.
 It is therefore only in relatively peripheral ion$-p$ collisions that the probability of a semi--exclusive production, in which we allow for low-mass proton dissociation but no other outgoing particles except for the intact heavy ion and the C-even meson in the central region, is large enough. We first study the dependence of this contribution on the number, $A$, of nucleons in the heavy ion.

\subsection{$A$ dependence of CEP${}^*$ cross section in $p$Pb collisions}

For simplicity we will consider the case of genuine exclusive production, that is where the proton remains intact. To allow for the possibility of $p\to X$ low-mass dissociation we simply  replace $\sigma^{nn}_{\rm CEP}$ by $\sigma^{nn}_{\rm CEP^*}$ in the results which follow. As we will discuss below, the transverse scale of the interaction amplitude between the proton and the nucleon in the lead ion is significantly smaller than the ion extent. This will remain a good approximation when the proton is allowed to dissociate, as here the transverse scale is very similar to the pure exclusive case. We note that the results presented below closely follow the arguments discussed in~\cite{Harland-Lang:2018iur}, and further details can be found there.

We denote the value of CEP cross section in nucleon-nucleon
 collisions by $\sigma^{\rm CEP}$ and the total cross section as
$\sigma_{\rm tot}$. 
For the nucleon density in the heavy ion we use the Woods--Saxon distribution~\cite{Woods:1954zz}
\be
\rho_N(r)= \frac{\rho_0}{1+\exp{(r-R)/d}}\;,
\ee
where $d$ characterises the skin thickness and $R$ the radius of the nucleon density in the heavy ion. For $^{208}$Pb
we take~\cite{Tarbert:2013jze,Jones:2014aoa}
\begin{align}\nonumber
R_p &= 6.680\, {\rm fm}\;, &d_p &= 0.447 \, {\rm fm}\;,\\ \label{eq:pbpar}
R_n &= (6.67\pm 0.03)\, {\rm fm}\;, &d_n &= (0.55 \pm 0.01) \, {\rm fm}\;.
\end{align}
The nucleon densities, $\rho$, are normalized to 
\be
 \int\rho_p(r)d^3r=Z \;, \qquad \int\rho_n(r)d^3r=N\;,
\ee
for which the corresponding proton (neutron) densities are $\rho_0 = 0.063$ (0.093) ${\rm fm}^{-3}$.

To study  those $p$Pb processes which lead to large rapidity gaps in the final state, there are two options.
 Either we have incoherent semi--exclusive production with
 rapidity gaps around the meson, where the ion dissociates, or coherent semi-exclusive production, where it remains intact. The former can be excluded experimentally by using the Zero Degree Calorimeter (ZDC) detectors, which can be used to veto on the neutral particles produced by ion dissociation. We consider both contributions below.

In the absence of survival effects, that is in the unrealistic limit of ignoring the impact of nucleon-nucleon  interactions in addition to the CEP process, the incoherent cross section is simply given by
\be\label{eq:sigincoh}
\sigma_{\rm incoh} = \int {\rm d}^2 b_{\perp}T_{A}(b_{\perp}) \sigma_{\rm CEP}^{nn} = A\cdot  \sigma_{\rm CEP}^{nn}  \;,
\ee
where we work in the approximation that the CEP interaction can be treated as point--like in comparison to the transverse extent of the ion.
That is, the cross section for incoherent interactions between the proton and nucleons in the ion, simply scales like $\sim A$, as we would na\"{i}vely expect. Here we have introduced the transverse nuclear density
\be\label{eq:tpn}
T_A(b_\perp)= \int {\rm d}z\, \rho_A(r) = \int {\rm d}z\, (\rho_n(r) + \rho_p(r))\;.
\ee
To calculate the cross section due to coherent interactions
 we must work at amplitude level. 
We fix the normalization of the CEP amplitude via the condition
\be
\sigma_{\rm CEP}^{nn} = \int {\rm d}^2 q_\perp | A_{\rm CEP}^{nn}(q_\perp)|^2\;,
\ee
where $q_\perp$ is the transverse momentum of the outgoing proton. 
Then the coherent production amplitude in proton--ion collisions is  given by
\be\label{eq:aceppa}
A_{\rm CEP}^{pA}(q_\perp) = A_{\rm CEP}^{nn}(q_\perp) F_A(Q^2)\;,
\ee
where the ion form factor is related to the nuclear density by
\be
F_A(Q^2) = \int {\rm d}^3r \, e^{i \vec{q}\cdot \vec{r}}  \rho_A(r)\;,
\ee
in the rest frame of the ion; in this case we have $\vec{q}^2 = -Q^2$. For our purposes we can simply take $Q^2 \sim q_\perp^2$. In impact parameter space \eqref{eq:aceppa} corresponds to
\begin{align}
\tilde{A}_{\rm CEP}^{pA}(b_\perp) &= \int {\rm d}^2 b_\perp' \tilde{A}_{\rm CEP}^{nn}(b_\perp') T_A(b_\perp -b_\perp')\;,\\
&\approx T_A(b_\perp) \int  {\rm d}^2 b_\perp' \tilde{A}_{\rm CEP}^{nn}(b_\perp') \;,\\
& = T_A(b_\perp)A_{\rm CEP}^{nn}(q_\perp = 0)\;,
\end{align}
where in the second line we have used the approximation that the size of the nucleon-nucleon amplitude is much less than the extent of the ion, i.e. $b_\perp ' \ll b_\perp$. The coherent cross section then becomes
\begin{align}
\sigma_{\rm coh} &= \int {\rm d}^2 q_\perp |A_{\rm CEP}^{pA}(q_\perp)|^2 \;,\\
&= 4\pi^2 \int {\rm d}^2 b_\perp |\tilde{A}_{\rm CEP}^{pA}(b_\perp)|^2\;,\\ 
 &\simeq 4\pi^2 |A_{\rm CEP}^{nn}(q_\perp = 0)|^2 \int {\rm d}^2 b_\perp T_A(b_\perp)^2\;,\\ \label{eq:tacoh}
 & =4\pi \frac{\sigma_{\rm CEP}^{nn}}{\left \langle q_\perp^2 \right\rangle} \int {\rm d}^2 b_\perp T_A(b_\perp)^2\;,
\end{align}
where we have defined
\be
\langle q_\perp^2 \rangle \equiv \sigma_{\rm CEP} /(\pi |A_{\rm CEP}^{nn}(q_\perp = 0)|^2).
\ee
When we use that $T_A(b_\perp) \sim R$, we find that the coherent cross section scales as $T^2R^2\propto A^{4/3}$.  This is to be expected: the short--range CEP interaction can only be coherent over the Lorentz--contracted $z$ direction, thus we only have a coherent enhancement $\sim (A^{1/3})^2$ in that direction, rather than the $\sim A^2$ scaling characteristic of a long--range interaction. 

We note that the above result holds for a CEP amplitude which is non--zero at $q_\perp=0$. However, for the production of a $0^-$ meson the amplitude will contain an antisymmetric tensor $\epsilon^{\alpha\beta\mu\nu}$ whose indices must be saturated by the transverse momentum $q_\perp$, and this will therefore vanish at $q_\perp=0$. In other words, we would have 
\be
\sigma_{\rm CEP, 0^-}^{nn} = \int {\rm d}^2 q_\perp q_\perp^2 | B_{\rm CEP}^{nn, 0^-}(q_\perp)|^2\;,
\ee
where $B_{\rm CEP}^{nn, 0^-}(q_\perp)$ is non--vanishing as $q_\perp \to 0$. The amplitude $ \vec{A}_{\rm CEP}^{nn, 0^-}(q_\perp)~=~\vec{q}_\perp  B_{\rm CEP}^{nn, 0^-}(q_\perp)$ can then be included as above via
\begin{align}
\tilde{\vec{A}}_{\rm CEP}^{pA}(b_\perp) &= \int {\rm d}^2 b_\perp' \frac{{\rm d} \tilde{B}_{\rm CEP}
^{nn,0^-}(b_\perp')} {{\rm d} \vec{b}_\perp'}T_A(b_\perp -b_\perp')\;,\\
&=  -\int {\rm d}^2 b_\perp' \tilde{B}_{\rm CEP}
^{nn,0^-}(b_\perp')\frac{{\rm d} T_A(b_\perp -b_\perp')}{{\rm d} \vec{b}_\perp'}\;,\\
&\approx  \frac{{\rm d} T_A(b_\perp)}{{\rm d} \vec{b}_\perp}B_{\rm CEP}
^{nn,0^-}(q_\perp = 0)\;,
\end{align}
for which we have
\be \label{eq:codd}
\sigma_{\rm coh} = 4\pi \frac{\sigma_{\rm CEP}^{nn}}{\left \langle q_\perp^2 \right\rangle^2} \int {\rm d}^2 b_\perp\left|  \frac{{\rm d} T_A(b_\perp)}{{\rm d} \vec{b}_\perp}\right|^2\;,
\ee
where we define $\left \langle q_\perp^2 \right\rangle^2 \equiv \sigma_{\rm CEP} / (\pi |B_{\rm CEP}^{nn, 0^-}(q_\perp = 0)|^2)$.

\subsection{Including the gap survival factor}

In the above expressions we have omitted the impact of accompanying 
 nucleon-nucleon interactions to the one involved in the CEP process. These can lead to secondary particle production that will fill the rapidity gaps. The impact of this is significant, and as we will see has an important effect on the expected scaling of the CEP cross sections. To account for the survival factor $S^2$, that is the probability that the  gaps are not filled by secondary particle production, we must multiply the integrands \eqref{eq:sigincoh} and \eqref{eq:tacoh} by 
\be
\label{gap-exp}
S^2(b_\perp)= \exp\left(-\sigma_{\rm tot}^{nn}\int d^3r'\rho_A(r')\delta^{(2)}(\vec b_\perp-\vec b_\perp')\right)\;,
\ee
where $\sigma_{\rm tot}^{nn}$ is the total nucleon--nucleon cross section and $b'_\perp$ is the transverse component of $r'$; we take 
$\sigma_{\rm tot}^{nn}$ rather than $\sigma_{inel}$ as even in elastic scattering the relatively large momentum transfers involved will tend to lead to ion break up.

That is, we multiply by the Poissonian probability for no additional inelastic 
 nucleon-nucleon interactions, considering all possible nucleon positions, but with the restriction that the overall proton--ion impact parameter is fixed. In other words, we only integrate over the longitudinal component of $r$, allowing this to be simply written in terms of the ion transverse density as
\be
S^2(b_\perp)= \exp\left(-\sigma_{\rm tot}^{nn} T_A(b_\perp)\right)\;.
\label{eq:39}
\ee
Now, numerically the exponent of \eqref{gap-exp} is very large, i.e.
\be
\sigma_{\rm tot}^{nn} T_A(b_\perp) \gg 1 \qquad  {\rm for} \qquad r \lesssim R\;.
\ee
Thus, in this region the probability for no additional particle production is strongly (exponentially) suppressed, and the only possibility to have a non--negligible exclusive (or semi--exclusive) cross section is to be close to the ion periphery, where the nucleon density and hence inelastic interaction probability is lower. We therefore expect that the scaling of e.g. \eqref{eq:sigincoh}, where all $A$ nucleons in the ion can participate equally in the CEP interaction, will significantly overestimate the predicted cross section. 
To be concrete, in the $r\gtrsim R$ region the nuclear density may be written as
\be
\rho(r)\approx  \rho_0 \exp(-(r-R)/d)\;.
\ee
Then, as shown in detail in~\cite{Harland-Lang:2018iur}, the CEP cross section simply comes from a ring of radius $\sim R$ and width $\sim d$. In particular, for the incoherent cross section we have
\begin{align}\label{eq:sigincohs2}
\sigma_{\rm incoh} &= \sigma_{\rm CEP}^{nn} \int {\rm d}^2 b_{\perp}T_{A}(b_{\perp}) e^{-\sigma_{\rm tot}^{nn} T_A(b_\perp)}\;,\\
&\approx \sigma_{\rm CEP}^{nn} \cdot 2\pi R\int  {\rm d} \delta x\, T_{A}(b_{\perp}) e^{-\sigma_{\rm tot}^{nn} T_A(b_\perp)}\;,\\
&\sim \sigma_{\rm CEP}^{nn} \cdot 2\pi R d\;,\\
& \sim  \sigma_{\rm CEP}^{nn} \cdot A^{1/3} \;,
\end{align}
where in the second step we integrate over the transverse displacement $\delta x$ with respect to the ion radius $R$ in the direction of the proton--ion impact parameter. Thus we expect the incoherent cross section to scale as $A^{1/3}$, rather than the na\"{i}ve $A$ scaling (\ref{eq:sigincoh}) of above. 

For the coherent case, the expected scaling is the same, however the factor of $\left \langle q_\perp^2\right \rangle$ in \eqref{eq:tacoh} in principle leads to some additional numerical suppression. In particular we find that~\cite{Harland-Lang:2018iur}
\be\label{eq:qtsub}
\frac{4\pi}{\sigma_{\rm tot}^{nn} \left \langle q_{\perp}^2 \right \rangle} \sim \frac{4\pi}{90 \,{\rm mb}  \cdot 0.1 \,{\rm GeV}^2} \sim 0.5\;,
\ee
where we as discussed in~\cite{Harland-Lang:2018iur} we take a rather small value of $\left\langle q_{\perp}^2 \right \rangle \sim 0.1$ ${\rm GeV}^2$ to account for the impact of the nucleon--nucleon survival factor on the average $q_\perp$ in the CEP cross section. For the production of a $0^{-}$ state, the logic is similar, but we instead apply \eqref{eq:codd}. The derivative picks up a factor of $1/d^2$, giving an overall factor of $1/(\left \langle q_\perp^2 \right\rangle d^2) \sim 1$ in comparison to the $0^{+}$ case, and so we expect no significant additional suppression in this case. We note however that this only corresponds to an approximate estimate, and in particular the value of $ \left \langle q_\perp^2 \right\rangle$ is not universal between the even and odd parity cases. More precisely, in the odd parity case the amplitude $A^{0^-}$ vanishes at
   $q_\perp=0$ and so the corresponding mean value of $\left\langle q^2_\perp \right \rangle$ is expected to be larger. Indeed, as we will see below, in the precise numerical calculation we do find some additional suppression in the odd parity case.

Recalling that the photoproduction cross section scales as $Z^2$, we can see that in the QCD--initiated case the expected scaling $\sim A^{1/3}$ is much milder. Thus this background, and in particular the coherent contribution, which is further numerically suppressed, may be under control. We will quantify these statements below.

\subsection{Double counting and the core effect}

However, before doing so there is one further complication to consider. In particular, the survival factor given by \eqref{gap-exp} and \eqref{eq:39} includes the possibility of additional interactions with any of the nucleons in the ion, but in the case of the active nucleon which undergoes CEP, this is already included in the value of $\sigma^{nn}_{\rm CEP}$. We must therefore exclude the contribution from this active nucleon in the optical density $T_A(b_\perp)$ in \eqref{eq:39}. As discussed in more detail in~\cite{Harland-Lang:2018iur}, this can be achieved by a suitable correction, removing the nucleon from the contributing outer shell of the ion and accounting for the repulsion between the nucleons by removing an interval $\pm r_{\rm core}$ around the active nucleon from the calculation of $T_A(b_\perp)$. Here $r_{\rm core}=0.6-0.8$ fm corresponds to the `core' region where there is a strong repulsive potential between the active nucleon and surrounding nucleons, the size of which can be estimated from the well understood deuteron case~\cite{Reid:1968sq}.
 
Excluding the interval $2r_{\rm core}\sim 1.4$ fm  we find roughly a 30\% correction in the power of the exponent, $\nu=\sigma^{nn}_{\rm tot}~T_A(b_{\perp})$. As the dominant contribution to the cross section comes from the ion periphery, where $\nu\simeq 1$, the value of the survival factor will increase by about 30~-~50\%. We will consider $r_{\rm core}=0.6-0.8$ fm in the results below as a reasonable range of values, in the absence of a more precise description of the nucleon-nucleon correlations in the heavy-ion periphery. 

\section{Background from Radiative Decays}\label{sec:bgrad}

\begin{figure} [htb]
\begin{center}
\vspace*{-0.0cm}
\includegraphics[scale=0.6]{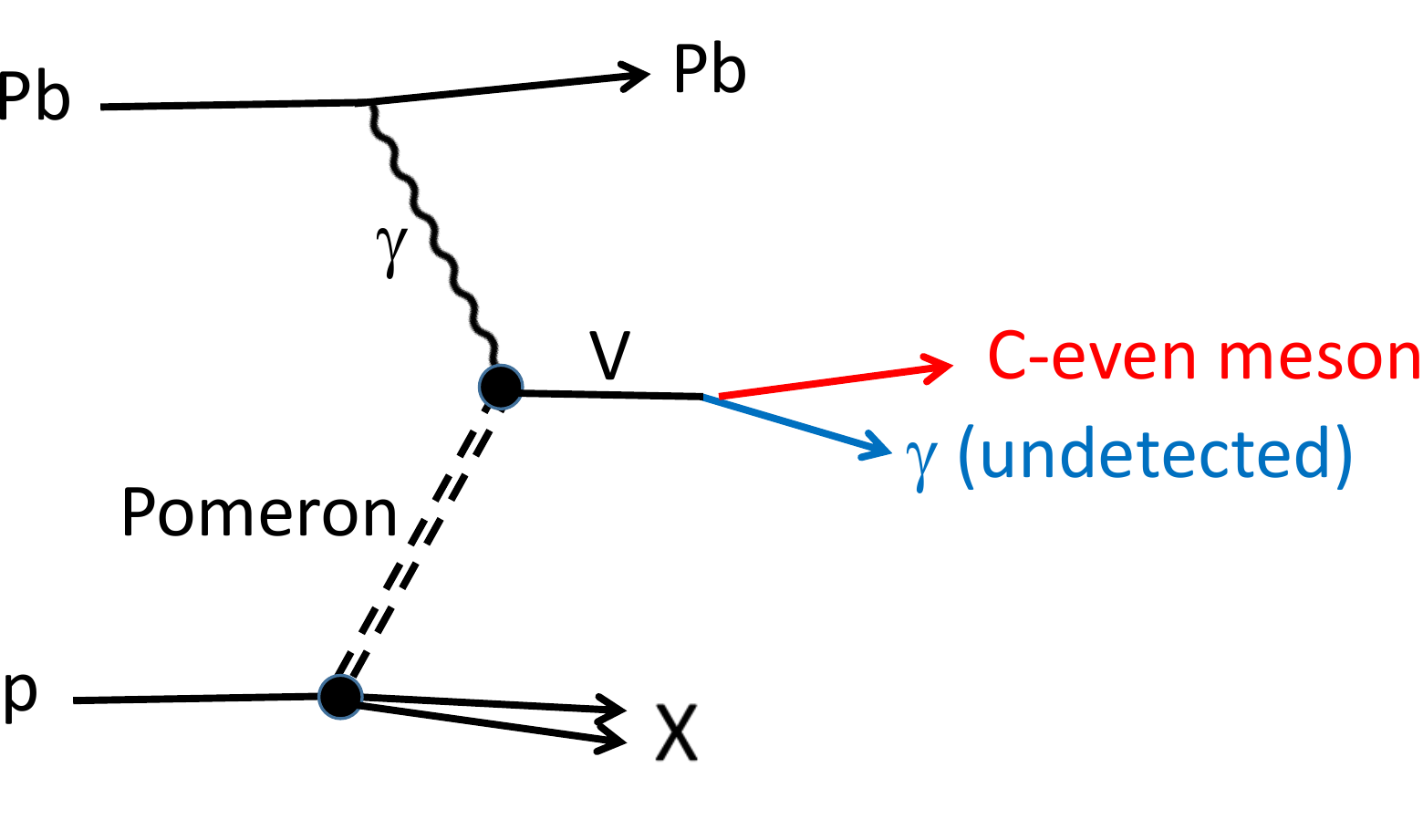}
\caption{\sf A background due to the exclusive production of C-even mesons via vector meson production followed by their radiative decay where the emitted photon is not detected.}
\label{fig:4}
\end{center}
\end{figure}

We in general also have to consider the background from the photoproduction of vector mesons ($ J/\psi,~\phi,~\rho...$), followed by their radiative decay to the signal $C$--meson, see Fig.~\ref{fig:4}. 
While not an irreducible background to the Odderon signal, experimentally this may generate a contribution, as the decay photon will not always be detected. We consider this possibility explicitly in the following sections.

\section{Numerical Predictions}\label{sec:num}

\begin{table}
\begin{center}
\begin{tabular}{|c|c|c|}
\hline
&exclusive& semi--exclusive\\
\hline
$r_{\rm core}$ (fm) & ~~0.8~~~~~~0.6~~ &~~0.8~~~~~~0.6~~ \\
\hline
P even&~2.5~~~~~~3.0~~&4.7~~~~~~5.3\\
P  odd &~0.45~~~~~0.5~~& ~4.7~~~~~~5.3~~\\
   \hline  
\end{tabular}
\caption{\sf The ratio $\sigma_{p{\rm Pb}}/\sigma_{pp}$ for exclusive and semi--exclusive CEP of odd ($0^-$) and even ($0^+$, $2^+$) parity mesons, produced through Pomeron-Pomeron fusion at $\sqrt{s_{nn}}=5.02$ TeV, shown for two values of $r_{\rm core}$.}
 \label{tab:fscal}
\end{center}
\end{table}

Table~\ref{tab:fscal} shows the ratio of the $p$Pb to $pp$ CEP cross sections for odd and even parity mesons produced through Pomeron-Pomeron fusion, calculated using the complete treatment described in detail in~\cite{Harland-Lang:2018iur}. To evaluate the  backgrounds to the Odderon contribution caused by Pomeron exchange in $p$Pb collisions we then need to input the corresponding cross sections in $pp$ collisions. Theoretically, these are not straightforward to 
  predict precisely, and so it is more reliable to take the existing experimental measurements of these $pp$ processes as a guide. We will for concreteness quote all results for central production in the $|Y_M|<1$ rapidity region, that is integrated over 2 units of the meson rapidity in the central region. Although we do not consider it explicitly here, this should roughly correspond to the expectations for the LHCb detector, i.e. with $2 < Y_M <4.5$.

\subsection{$f_2$ production}

For the case of $f_2$ production, the semi--exclusive process (i.e. including an admixture due to proton dissociation)
 has been measured by CMS~\cite{Khachatryan:2017xsi} at 7 TeV. Using this information we can, to a good approximation, estimate the rate at 8.16 TeV collision energy relevant for $p$Pb collisions. We find for $|Y_M|<1$ that
\be
\sigma_{pp}^{f_2} = 2-3 \,\mu{\rm b}\;.
\ee
From Table~\ref{tab:fscal} we can see that his corresponds to roughly
\be
\sigma^{f_2,I\!P}_{p{\rm Pb}} = 5- 9 \,\mu{\rm b}\;,\qquad \sigma^{f_2,I\!P}_{p{\rm Pb}^*} = 9- 16 \,\mu{\rm b}\;,
\ee
where Pb$^*$ indicates the semi-exclusive (incoherent) cross section, i.e. where the lead ion does not remain intact. We can see from Table~\ref{tab:oddcs} that the expected upper limit to the Odderon-initiated cross section is
\be
\sigma^{f_2,{\rm O}}_{p{\rm Pb}} < 6.0  \,\mu{\rm b}\;,
\ee
at the 95\% confidence level, where we have integrated over the interval  $|Y_M| <1$.  Thus, unfortunately even considering only the purely exclusive case (i.e. with vetoes applied on the lead ion dissociation) the predicted background is somewhat larger than the experimental upper limit, and so this will be a very challenging channel in which to observe Odderon exchange.

The background due to vector meson radiative decay is rather small. The main contribution comes from the $J/\psi\to \gamma f_2$ decay, with branching ratio $1.4\cdot  10^{-3}$~\cite{Patrignani:2016xqp}. Accounting for the $J/\psi$ photoproduction cross section~\cite{Voss:2003iu} this gives a background in $p$-Pb collisions of only $\sim $ 0.02 $\mu$b. 

 \subsection{$\pi^0$ production}
 
 The $\pi^0$ meson cannot be produced by Pomeron-Pomeron fusion due to isospin conservation. This background will therefore be absent, and from Table~\ref{tab:oddcs} we see that the expected upper limit on the Odderon-initiated cross section is
\be
\sigma^{\pi^0,{\rm O}}_{p{\rm Pb}} < 14.8  \,\mu{\rm b}\;,
\label{eq:50}
\ee
for $|Y_M|<1$.
In principle the situation  looks
much more favourable in this case,
although, in practice, the observation of an exclusive $\pi^0$ signal could
be difficult for various experimental reasons. For example, the trigger
conditions for the detection of soft photons are challenging
and the presence of additional $\pi^0$ mesons in the final state in the
CEP${}^*$ process could complicate the signal. We note that as the pion cannot be produced via Pomeron-Pomeron fusion, the contribution from events with ion dissociation should already be small, and so there is no further benefit in applying a cut using the Zero Degree Calorimeter.

In addition to this there is a large background from $\omega$ meson production, followed by the radiative $\omega\to\gamma\pi^0$ decay. Taking a $\omega$ photoproduction cross section of about 2$\mu$b~\cite{Voss:2003iu} and a branching ratio of 8\%~\cite{Patrignani:2016xqp}, we expect a total background cross section of 60$\mu$b, that is roughly four times as large as the expected upper limit (\ref{eq:50}) of the signal. On the other hand, the background from ultraperipheral $\omega$ production where the additional photon from the radiative decay is detected can be measured and this background subtracted in a reasonably data--driven way. However, the overall size of the background suggests this will need a relatively careful treatment.
 
  \subsection{$\eta$ production}
 For the $\eta(548)$ meson, the dominant $SU(3)$--flavour octet component will, as in the $\pi^0$ case, not couple to the double Pomeron exchange production channel. This background will not be entirely absent, due to the non--zero $SU(3)$--flavour singlet component of the $\eta$, however this contribution will be suppressed by the relatively small mixing angle~\cite{Patrignani:2016xqp}
 \be
 \sigma^{\eta,I\!P}_{p{\rm Pb}} \propto \sin^2(\theta_1) ~=~ 0.03-0.1\;,
 \ee
 that is, by at least an order of magnitude in comparison to $f_2$ production.  
 However, in addition to this, we expect a further suppression due to the  the $J^P=0^-$ nature of the $\eta$ meson, the production of which is typically suppressed in double Pomeron exchange~\cite{Kaidalov:2003fw}. In particular, the $\eta$ production vertex must be symmetric with respect to the two Pomeron exchanges and simultaneously contain an antisymmetric tensor, due to the odd parity of the meson. The simplest expression satisfying these requirements is
 \begin{equation}
 \label{vert}
V \propto \epsilon_{\alpha\beta\mu\nu}k_{1L}^\alpha k_{2L}^\beta k_{1\perp}^\mu k_{2\perp}^{\nu}\ ,
 \end{equation}
 where $k_{1t},k_{2t}$ and $k_{1L}, k_{2L}$ are the transverse and longitudinal momenta exchanged through the Pomerons.
 We therefore have
  \be
\frac{ \sigma^{\eta,I\!P}_{p{\rm Pb}}}{{\rm d}^2k_{1\perp}{\rm d}^2k_{2t}}\propto k^2_{1\perp}k^2_{2\perp}\;,
 \ee
 where these factors of $k_\perp^2$ are absent in the even parity case.  The dimension of these factors must be compensated by some additional scale. While we do not know the precise value of this scale for such a non-perturbative process, we can expect it driven by the Pomeron size, that is by the value of $\alpha'_{I\!P}$, i.e. the slope of Pomeron trajectory. Phenomenologically,  $\alpha'_{I\!P}\sim 0.25 $ ${\rm GeV}^2$ is rather small, and so we expect
 \be
  \sigma^{\eta,I\!P}_{p{\rm Pb}} \propto \sin^2(\theta_1)  (\left\langle k_\perp^2\right \rangle \alpha'_{I\!P})^2 \lesssim 10^{-3}\;,
 \ee
 where we have taken the rather conservative values of $\theta_1\sim 20^\circ$ and $\left\langle k_\perp^2\right \rangle =0.5$ ${\rm GeV}^2$; that is, the suppression may be significantly stronger. For example, if we take the value of $\theta_1\sim 10^\circ$ preferred by~\cite{Feldmann:1998vh}, then the suppression is 
 about 4 times larger. Taking the observed $f_2$ cross section as a baseline, we therefore roughly expect the background due to double Pomeron exchange to be $O(0.01)$ $\mu$b, and potentially much smaller than this. From Table~\ref{tab:oddcs} we see that this is significantly less than the predicted upper limit on the Odderon-initiated cross section (for $|Y_M|<1$) of
\be
\sigma^{\eta,O}_{pPb} \lesssim 6.8  \,\mu{\rm b}\;.
\ee
 Finally, we have to consider the background due to radiative vector meson decay, in particular from $\phi\to\gamma\eta$ when the relatively soft additional photon escapes detection. With $\sigma(\gamma p\to\phi+p)=1\ \mu$b~\cite{Voss:2003iu} and 1.3\% branching~\cite{Patrignani:2016xqp} we expect a total background cross section of about  5 $\mu$b, that is smaller than but comparable to the upper limit on the signal.
 We in addition expect a total background cross section of $\sim$ 1.2 $\mu$b from  $\rho\to\gamma\eta$ decay. In both cases by measuring the corresponding backgrounds where the additional photons are detected, we should be able to subtract these, but again some care is clearly needed. 
 Finally there is in principle a contribution coming from
$\eta'$ production in  Pomeron-Pomeron fusion, followed by the $\eta'\to\eta\pi^0\pi^0$ decay.
 However, we would expect this to be efficiently rejected by observing at least one of four photons from $\pi^0$ decay in the final state.
 
 \subsection{$\eta_c$ production}
Theoretically, $\eta_c$ photoproduction has the advantage that due to the relatively large charm quark mass, $m_c$, the perturbative QCD calculation of the Odderon--meson coupling is more justified, although the overall amplitude is nonetheless sensitive to the infrared region due to the modelling of the Odderon--proton coupling.
 On the other hand, due to the larger meson mass the cross section is much smaller. In~\cite{Bartels:2003zu,Czyzewski:1996bv,Ma:2003py,Engel:1997cga} this was estimated to be $\sigma(\gamma p\to \eta_c+X)\sim 60$ pb in total, and about 10 pb for $|t|>1$ GeV$^2$. That is we expect
 \begin{equation}
 \frac{d\sigma}{dy}=2-10\,{\rm nb}\;,
\end{equation}  
in $pPb$ collisions. The background from Pomeron--Pomeron fusion was calculated in~\cite{Harland-Lang:2014lxa} to be roughly
\be
\sigma_{pp}^{\eta_c} = 0.4\,{\rm nb}\;,
\ee
at $\sqrt s=7-14$ TeV, corresponding to 
\be
\sigma^{\eta_c,I\!P}_{p{\rm Pb}} = 0.2 \,{\rm nb}\;,\qquad \sigma^{\eta_c,I\!P}_{p{\rm Pb}^*} = 2 \,{\rm nb}\;,
\ee
where the uncertainty in the pQCD prediction is larger ($\sim {}^{\times}_{\div} 2$) than the spread in Table~\ref{tab:fscal} and we therefore only quote the approximate central values for illustration. This is therefore smaller than the expected Odderon signal, in particular for the case of no ion dissociation. Clearly a measurement of $\eta_c$ production in the $pp$ mode would further help calibrate this background.

Unfortunately, the branching ratios of accessible $\eta_c$ decay channels are all rather small. The most convenient modes for experimental observation are~\cite{Patrignani:2016xqp}:
\be
\eta_c\to \rho\rho\;: \qquad (1.8\pm 0.5) \%\;,
\ee
where the branching ratio is given. This corresponds to 0.6\% in the $2\rho^0\to 2(\pi^+\pi^-)$ mode. In addition we have
\be
\eta_c\to K^{*0}K^-\pi^+\;: \qquad (2.0\pm 0.7) \%\;,
\ee
 giving about 1.3\% in the $K^-K^+\pi^+\pi^-$ mode, and 
 \be
 \eta_c\to K\bar K\pi\;: \qquad (7.3\pm 0.5) \%\;,
 \ee
 which can be observed via the $K^0_s\pi^+K^-$ and $K^0_s\pi^-K^+$ channels for which we get a $\simeq 1.7\%$ branching. Finally we have 
 \be
 \eta_c\to 3(\pi^+\pi^-) \;: \qquad (1.8\pm 0.4) \%\;.
 \ee
 Summing over all of the charged decay modes, we find a total branching ratio of $\sim 5\%$, corresponding to an expected Odderon signal of $\sim 0.1-0.5$ nb. 
The cross section alone is certainly large enough to be experimentally feasible, however it is very challenging to separate such a signal from minimum bias events, where the cross section ($\sim  1$ b) is more than a factor $\sim 10^9$ larger. To bypass this, it would therefore be essential to select ultraperipheral events with Large Rapidity Gaps at the trigger stage.

A potentially even more serious issue arises due to the background from $J/\psi\to\gamma\eta_c$ decay, which occurs with branching $1.7\%$~\cite{Patrignani:2016xqp}.
This corresponds to a total background cross section of $\sim 250$ nb, excluding the $\eta_c$ branchings, that is more than 20 times larger than the expected Odderon induced cross section. Clearly accounting for such a background will be highly challenging.

\subsection{Summary of numerical predictions}
 
 We draw together in Table \ref{Tab3} the results for the Odderon signal and backgrounds for the various C-even mesons, $M$, discussed above. The first column shows the upper limit of the cross section for the Odderon-exchange process (Pb $p \to $Pb$~M~X$) deduced from the HERA data for ($\gamma ~p\to M~X$) as described in Section \ref{sec:gen} and listed in Table \ref{Tab1}.  The other columns give the cross sections expected for the Odderon signal and background in the (semi) exclusive process
 \be
 {\rm Pb} ~ p~\to  ~{\rm Pb}+M+X,
 \ee
 where here the $+$ signs denote the presence of large rapidity gaps. Note the Table shows $d\sigma/dY_M$ at $Y_M=0$ while the text presents numbers for $\sigma$ integrated over the central interval $|Y_M|<1$, so the cross section is essentially twice as large. 
 \begin{table}
\begin{center}
\begin{tabular}{|c||c|c||c|c|c||} 
  \hline
 C-even  & \multicolumn{2}{c||}{Odderon Signal} & \multicolumn{3}{c||}{Backgrounds}\\ 
\cline{2-6} 
 meson ($M$)& Upper & QCD  &  & Pomeron- & \\ 
 & Limit & Prediction &$\gamma\gamma$& Pomeron &$V\to M+\gamma$ \\
 \hline\hline
$\pi^0$ & 7.4 & 0.1~-~1 & 0.044 & -- & 30  \\ \hline
$f_2(1270)$ & 3 & 0.05~-~0.5 & 0.020 & 3~-~4.5 & 0.02 \\ \hline
$\eta(548)$ & 3.4 & 0.05~-~0.5 & 0.042 & {\rm negligible} & 3 \\ \hline
$\eta_c$ & -- & $ (0.1-0.5)\cdot 10^{-3}$ & 0.0025 & $\sim 10^{-5}$ & 0.012\\ \hline
\end{tabular}
\caption{\sf The expected cross sections 
 ($d\sigma/dY_M$ at $Y_M=0$ in $\mu$b) of the Odderon signal and backgrounds in the CEP* ultraperipheral production of C-even mesons $(M)$ in high-energy proton-lead collisions ($Pb+p \to Pb+M+X$) integrated over the interval $0.2<p_\perp<1$ GeV. In the $\eta_c$ case a total branching ratio of 0.05 has been applied, i.e. summing over the channels discussed in the text.}
\label{Tab3}
\end{center}
\end{table}  
 
 \section{Conclusions and Outlook}\label{sec:conc}
 
 In this paper we have discussed the possibility of observing Odderon exchange in ultraperipheral $pPb$ collisions. We have shown that the signal cross sections for the semi--exclusive production of $C$--even light mesons due to Odderon exchange could be quite large, up to the $\mu$b level, and so represent a viable search channel. However, it is important to emphasise that an experimentally feasible signal cross section is a necessary but not sufficient condition when searching for the so--far rather elusive QCD Odderon. In particular, it is essential to consider the purity of the expected signal, and to quantitatively estimate the contribution from all potential backgrounds. This has been the main goal of this paper.
    
  We have in particular considered the two principle irreducible backgrounds, due to  photon--photon and Pomeron--Pomeron fusion. While in $AA$ collisions the contributions from the former, which is enhanced by $\sim Z^4$ will be overwhelming, in $pp$ collisions the contribution from the latter will be strongly dominant. We have therefore identified $pA$ collisions, specifically $p$-Pb in the context of the LHC, as the most promising channel. In this case, we have found that the the  photon--photon background can be strongly suppressed by requiring that an appropriate cut ($p_\perp \gtrsim 0.2$ GeV) is placed on the produced meson, while leaving the signal roughly unchanged. The background from Pomeron--Pomeron fusion is found to scale quite mildly with mass number $\sim A^{1/3}$, due to the suppression induced by requiring that no additional particles be produced. This leads the background, which naively we might expect to be overwhelming, to be of the same order or even smaller than the Odderon--exchange signal.

 We have considered several candidate $C$--even mesons with which to observe Odderon exchange: the $\pi^0$, $\eta(548)$, $\eta_c$  and $f_2(1270)$. For the $C$--even $f_2$ meson  there is very low background due to the vector meson decay (like $J/\psi\to\gamma f_2$) while the background due to double Pomeron exchange is found to be somewhat larger than any possible signal, in particular taking into account existing HERA upper  limits. In principle, the signal-to-Pomeron background ratio could be greatly improved by selecting events with very small outgoing ion transverse momenta, however this is not experimentally feasible in the present LHC experiments. On the other hand, the production of the $SU(3)$ flavour octer $\pi^0$ is forbidden in double Pomeron exchange, and so such a background is entirely absent, although this may prove to be an experimentally challenging channel. Moreover here we have a larger reducible background due to radiative $\omega$ decay, $\omega\to \pi^0\gamma$. For the $\eta(548)$, the relatively small flavour singlet component combined with the suppression expected due to the odd parity of the meson leads us to expect the Pomeron-Pomeron background to be many orders of magnitude below the Odderon--initiated signal. Unfortunately in this case the reducible contributions coming from $\phi\to\gamma\eta$ and $\eta'\to\eta\pi^0\pi^0$ decays are again rather large.
 Finally, the production of the heavier $\eta_c$, while in principle representing a viable channel, has a much smaller production cross section  and the signal will also be reduced by the small branching ratios of the experimentally accessible decay channels. The results are summarized in Table \ref{Tab3}.
 
 Our study is not intended to be exhaustive, and the production of heavier mesons such as the $\chi_c$, or light mesons such as the   $f_0(980)$,  $f_1(1285),f_1(1420),\eta(1405)$ and $\eta(1475)$, though not considered explicitly here, may also be worth exploring. In particular the $f_1(1285)$ has the advantage of a relatively small width and the fact that due to its $J^P=1^+$ quantum numbers the backgrounds due to Pomeron-Pomeron and/or $\gamma\gamma$ fusion should be suppressed.

To conclude, our study highlights the importance of considering the background contributions in the case of observing Odderon exchange via photoproduction in $p$-Pb collisions at the LHC. In particular, while we find that the expected signal is certainly sufficiently large to be observed at the LHC, these backgrounds are not always negligible and can pose a challenge. 
Here we have discussed various techniques for bringing the backgrounds under control, paving the way for future experimental searches at the LHC.
 
\section*{Acknowledgements}

We thank Ronan McNulty and E.L. Kryshen for the interest to this work and for useful discussions. We are grateful to Vadim Isakov for the discussion of the proton and neutron distributions in a heavy nucleus.  LHL thanks the Science and Technology Facilities Council (STFC) for support via grant awards ST/L000377/1 and ST/P004547/1. 
MGR thanks the IPPP at the University of Durham for hospitality. VAK acknowledges  support from a Royal Society of Edinburgh  Auber award.  

\bibliographystyle{JHEP.bst}
\bibliography{Odd2.bib}

\end{document}